\documentclass[aps,prl,superscriptaddress,twocolumn]{revtex4-1}

\usepackage{amsmath}
\usepackage{graphicx}
\usepackage{epstopdf}
\usepackage{bm}
\usepackage{subfigure}

\begin{document}


\title{Exclusive Quantum Channels in Quantum Networks}


\author{Xi Chen}
\author{He-Ming Wang}
\author{Liang-Zhu Mu}
\email{muliangzhu@pku.edu.cn}
\affiliation{School of Physics, Peking University, Beijing 100871, China}


\author{Heng Fan}
\email{hfan@iphy.ac.cn}
\affiliation{Institute of Physics, Chinese Academy of Sciences, Beijing 100190, China}
\affiliation{Collaborative Innovation Center of Quantum Matter, Beijing 100190, China}


\date{\today}

\begin{abstract}
Quantum state can be teleported to a remote site by only local measurement
and classical communication if the prior Einstein-Podolsky-Rosen quantum channel is available between
the sender and the receiver.
Those quantum channels shared by multiple nodes can constitute a quantum network.
Yet, studies on the efficiency of quantum communication between nodes of quantum networks remain limited,
which differs from classical case in that the quantum channel will be consumed if
teleportation is performed.
Here, we introduce the exclusive quantum channels (EQC) as the measure of
efficiency of quantum information transmission. It quantifies the amount of quantum information
which can be teleported between nodes in a quantum network.
We show that different types of EQC are local quantities with effective circles.
Significantly, capacity of quantum communication of quantum networks quantified by EQC
is independent of distance for the communicating nodes.
Thus, the quantum network can be dealt as the isotropic medium where quantum communication is no-decaying.
EQC are studied by both analytical and numerical methods.
The EQC can be enhanced by transformations of lattices of quantum network via entanglement swapping.
Our result opens the avenue in studying the quantum communication of the quantum networks.
\end{abstract}

\maketitle


Quantum information processing offers quantum algorithms which may surpass their classical
counterparts. Recently, quantum network and its extension, quantum internet which might
potentially be the next generation internet,
have been attracting a great deal of interests \cite{Kimble08,ChouCW,CiracZoller97,PappChoi,ChoiKimble10,Blattreview}.
The quantum networks
are constituted by quantum nodes where quantum information can be produced, stored and processed locally.
Those nodes are supposed to be linked by both quantum channels and classical channels
 so that quantum states can be
teleported between different nodes. This feature is also essential if each individual node
in quantum network is restricted to finite size of state space, and probably
the central quantum servers are available. The
quantum network may have the capabilities for quantum computation with a distributed style,
simulation of quantum many-body systems \cite{CuiNat,AmicoReview}, quantum metrology \cite{qmetro1},
quantum cloud computation with
unconditional security  by quantum key distribution protocols \cite{BarzKashefi}. These exciting features
provide the motivation to examine research related to quantum networks.

Quantum state can be sent directly between any nodes by flying qubits like photons in quantum networks.
On the other hand,
the prior shared entangled state between different nodes, which can be considered as the
available quantum channel, can facilitate quantum communication by only local operations on each individual
node with assistance of classical communication. This is the well-known teleportation \cite{PhysRevLett.70.1895}.
Additionally, the teleportation protocol may increase the fidelity for state transmission in case
entanglement purification is applied \cite{purify,DLCZ}.
This is a fundamental function of the quantum network.
The efficiency of such a quantum network for quantum communication should be well
described and quantified. One pioneering work is about the entanglement percolation
in quantum networks \cite{QEP}. It is shown that the problem of establishing maximally entangled
states between nodes can be related directly to classical percolation in statistical mechanics.
In particular, by applying appropriate local
measurements on some nodes which changes the lattice configurations of quantum network
and correspondingly changes the percolation threshold, the entanglement percolation
can change from failure to success. However, the entanglement percolation only describes
whether a quantum channel between two nodes can be constructed or not, i.e. it only
shows `on' or `off'.
If it succeeds, it is still not clear what amount of quantum information can be teleported.
In comparison with classical networks,
this is crucial for quantum networks since once the pre-shared entanglement is consumed
for telepotation, new quantum channel should be constructed.
Thus the number of exclusive quantum channels, i.e., independent maximally entangled pairs,
become important for quantum networks.

In this work, we define the EQC as the expectation number
of maximally entangled pairs, which equals to the number of individual qubits capable
teleported between two nodes in an established quantum network,
normalized by the number of bonds connecting each node.
We find that two types of EQC depend on the local structure and the bond singlet conversion probability
of the quantum network.
Particularly, they are independent of the distance of the studied nodes when they are
located farther than their effective radii. They can then be easily computed without considering
the detailed connections between these nodes. Significantly, it also means that
the EQC are non-exponentially decaying, thus avoiding the main hurdle for long-distance
quantum communication. The EQC can be increased by lattices transformation realized by quantum measurements
on some nodes.
We show that this quantity defines effectively the efficiency of the quantum communication for the quantum network.
The EQC are universal quantities applicable in quantum networks with various geometries.

\section{Exclusive quantum channels}
\begin{figure}
\includegraphics[width=84mm]{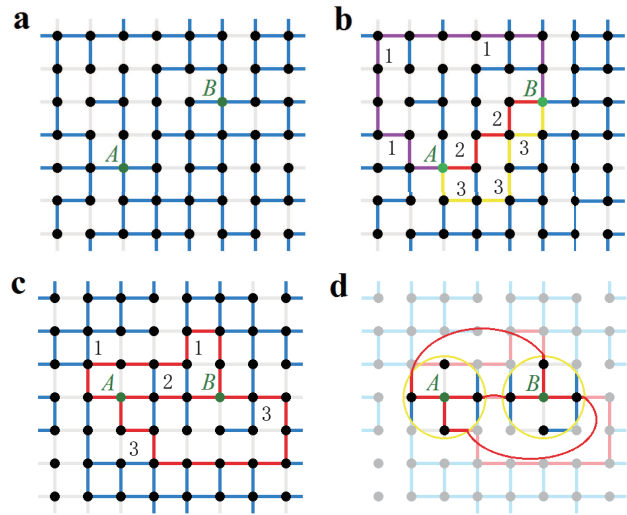}
\caption{A part of an infinite square network with $p=2/3$ after the singlet conversion. \textbf{(a)} Blue bonds are states that has been successfully converted to maximally entangled states while gray bonds represent those failed.
\textbf{(b)} The three independent quantum connections between A (Alice) and B (Bob) marked as green points are highlighted in red, yellow and purple (also 1, p2 and 3). \textbf{(c)} A part of an infinite square network with $p=2/3$ and $d=3$. The three quantum connections between Alice and Bob (green points) are highlighted in red (also 1, 2 and 3). \textbf{(d)} The attractive circles in the network. Each circle has 4 `internal bonds'. Outside the effective circles the network can be regarded simply as an isotropic `medium'.
So, whether a quantum channel can be formed or not mostly depends on how many `internal bond' is formed.}
\label{EQC}
\end{figure}

We consider a quantum network which constitutes a periodic two-dimensional lattice, see FIG. p\ref{EQC}.
The nodes are located at lattice sites, and each bond connecting nearest neighboring sites
correspond a pure entangled state shared by two nodes, $|\varphi\rangle=\sqrt{\lambda_1}|00\rangle+\sqrt{\lambda_2}|11\rangle$, where $\lambda_1+\lambda_2=1$ and $\lambda_1\geq\lambda_2$ is assumed.

Suppose we need to transmit some qubits from Alice at node $A$ in the network to Bob at node $B$.
We follow the standard protocol: First, we convert each entangled state
represented by a bond to a maximally entangled state.
The optimal conversion probability is $p={\rm min}(1,2(1-\lambda _1))$, \cite{vidal}.
The quantum network then is mapped to a classical network with each bond
appearing probability $p$, see Ref. \cite{QEP} and FIG. \ref{EQC} (a).
Secondly, if the bonds of maximally entangled states can form a path connecting $A$ and $B$,
by entanglement swapping (entanglement repeating), a maximally entangled state
can be generated, which constitutes a quantum channel, between $A$ and $B$.
Then Alice can teleport a qubit to Bob. The entanglement percolation describes well
whether this protocol can succeed or not \cite{QEP}.
However, different from classical case, the quantum channel can only be used once.
After a qubit is teleported through a quantum channel, the resource of
entanglement is consumed. The bonds no longer carry the entanglement.
So the number of EQC of a quantum network is a key
quantity in describing the efficiency of quantum communication, see FIG. \ref{EQC}.

We define the EQC as,
\begin{equation}
\mathrm{EQC}=\frac{N(p)}{N(1)}
\end{equation}
where $N(p)$ is the expectation number of quantum channels connecting $A$ and $B$, and $N(1)=N(p)|_{p=1}$
which is actually the number of bonds connecting each node in one to one communication
and thus is fixed for a lattice.
By definition, EQC increases with $p$, is upper bounded by 1, and depends on the network and the protocol used. Greater EQC means higher efficiency of qubit transmission.
It is generally very hard to find EQC analytically. However, as we can see below, the regular structures of lattices allow us to build a rather simple model which can approximate EQC well.


\section{Attractive and Repulsive Radii}
Now, we consider a quantum network with square lattice configuration.
We remark that our method is general and applicable for various quantum networks.
We also assume that $p>0.6$, which is clear above the percolation threshold
of square lattices, $p_{critical}=0.5$. When $p$ is near the critical point,
the quantum state transferring efficiency can be reasonably assumed to be low.
The effective approximation method may not be applicable,
so we will not discuss this situation.

\begin{figure}
\includegraphics[width=87mm]{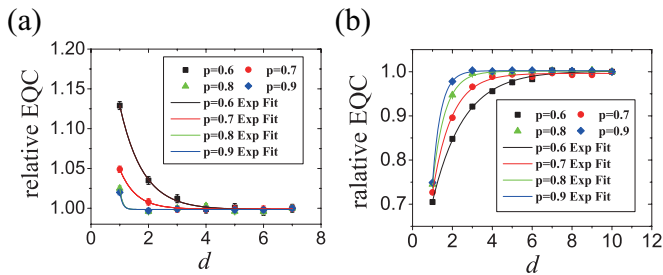}
\caption{\textbf{(a)} Monte Carlo estimations of EQC between two nodes in a square lattice for different $p$s. Since the attractive radii pertain to relative change of EQC, we depict the relative EQC in which all EQC of $d=7$ are normalized as 1. For the original image, see \cite{appendix} for details. The corresponding $\gamma$s for $p=0.6, 0.7, 0.8, 0.9$ are 1.2, 1.7, 10.6, 10.9, with $r_{at}=1.33, 1.09, 0.59, 0.59$.
A rather fast convergence can be observed when $p>0.8$. \textbf{(b)} Monte Carlo estimations of EQC from two nodes to infinity for different $p$s. For convenience in demonstration of repulsive radii we still present the relative EQC. All EQC of $d=10$ are normalized as 1.
For the original image, see \cite{appendix} for details. The corresponding $\gamma$s for $p=0.6, 0.7, 0.8, 0.9$ are 0.64, 1.04, 1.59, 2.37, with $r_{rp}=2.06, 1.46, 1.13, 0.92$.}
\label{MC}
\end{figure}
Let Alice and Bob be on two lattice nodes with distance $d$,
which is defined as the linear distance and
the distance of two nearest neighboring sites is assumed to be 1. Now Alice would like to send some qubits to Bob via the network.
Monte Carlo estimations (FIG. \ref{MC}) shows that when $p>0.6$, EQC between Alice and Bob decreases and converges quickly to its asymptotic value when $d$ increases. This fast convergence suggests us to define the attractive radius of an effective circle,

The EQC can be fitted by an exponential function,
\begin{equation}
  EQC=E_0+C_0e^{-\gamma_{at} d},
  \label{fitfunc}
\end{equation}
where $E_0$, $C_0$ are parameters depending on $p$. We then can relate the attractive radius with an exponent parameter, $r_{at}=1/\gamma_{at}+1/2$. When $d>2r_{at}$, i.e. the attractive circles are not overlap, the second term
in Eq.\ (\ref{fitfunc}) can be neglected since it is less than $e^{-2}$
as we generally used physically. Thus, we know that $E_0$ represents $EQC$
asymptotically when distance $d$ is large, this is what we are interested.
We will present an estimation of $E_0$ below.
This formula agrees well with numerical results, see FIG.\ref{MC}.
For $p>0.6$, we have a rough estimation for the attractive radius, $1<r_{at}<\sqrt{2}$.
This means that the effective circle of Alice or Bob includes only their nearest neighboring sites
on a square lattice.

The physical interpretation of $r_{at}$ can be understood as the radius of an effective circle.
When the distance of $A$ and $B$ is larger than $2r_{at}\approx 2.8$,
the EQC will be approximated being independent of their distance,
$EQC=E_0$ which is a fixed value for a lattice.
It seems like that the two nodes
are connected by first connecting from their positions to
the borders of their effective circle, and second the two circles are connected through
the isotropic medium. This result is significant since it means that the number of qubits
capable transmitted through a quantum network does not decrease with the increase of the distance.
It avoids the usual exponential decaying probability for long-distance quantum communication.
Our result is also consistent with the conclusion that quantum percolation in quantum networks can succeed
when $p$ is larger than a threshold. The EQC further quantifies the number of quantum channels
available in a quantum network.
When $A$ and $B$ are located near, it is possible that more quantum
channels can be constructed. So the attraction will be beneficial
which stimulates the name `attractive radius'.

Next, we consider a different scheme for quantum communication in quantum networks.
The scheme is that Alice and Bob at different nodes would like to send qubits to someone
at positions of infinity in the network. Here we only concern about how many qubits can be
sent out by Alice and Bob. A simple consideration is that if Alice and Bob are located near
in the quantum network, their ability of sending out qubits will be influenced by each other.
The problem is in what distance they can send out qubits independently.
We name the radius of this circle as the repulsive radius.
Monte Carlo estimations (FIG. \ref{MC}) show that when $p>0.6$,
EQC for Alice and Bob together to transfer states to
someone at infinity increases and converges quickly to its asymptotic value when $d$ increases,
similar as the case of the attractive circles.
Also, we fit the Monte Carlo data to an exponential function like Eq.\ (\ref{fitfunc}).
We find that EQC of this type is also described well by an exponential function (Fig. \ref{MC}),
\begin{equation}
  EQC=E'_0+C'_0e^{-\gamma_{rp} d},
\end{equation}
where $E'_0,C'_0$ are parameters similar to those in previous consideration of attractive radii.
Also we define the repulsive radius as, $r_{rp}=1/\gamma_{rp}+1/2$.
We emphasize that the EQC here means that Alice and Bob both send qubits to infinity.
For $p>0.6$, we have a rough estimation, $\sqrt{2}<r_{rp}<2$. It means that when distance of
Alice and Bob is larger than $2r_{rep}\approx 4$, they can be considered independent in sending
out qubits.

The repulsive radius $r_{rp}$ also has a clear physical interpretation.
It describes the range of the bonds that have a significant influence to the quantum channels
connecting A and B to infinity. For small $d$,
the paths which connect $A$ and $B$ with positions of infinity used to build the quantum channels
will conflict with each other, which will decrease the number of quantum channels resulting in
a smaller EQC.

The key feature of the EQC as a function of distance is its rapid convergence to a stationary value.
It is worth mentioning that this is a universal feature for various lattices.

\section{Evaluation of exclusive quantum channels}

Now we can easily approximate EQC analytically by using the physical meanings
of $r_{at}$ and $r_{rp}$. Suppose Alice needs to transmit qubits to Bob.
Two situations are considered: (i) Point-to-point quantum communication,
Alice and Bob both can access only one node in the network;
(ii) Multipartite-to-multipartite quantum communication,
they each can access $k$ nodes. Here we assume that each node
is independent from other nodes. That is, any pair distance of
those $k$ nodes possessed by Alice (Bob) is larger than  2$r_{rp}$.
And Alice's group nodes are far from Bob's group nodes.
Within this situation, our approach gives $E_0$ in Eq.\ (\ref{fitfunc}).
Henceforth, all EQC we mentioned here are $E_0$ since
distance is supposed large.

\begin{figure}
\includegraphics[width=88mm]{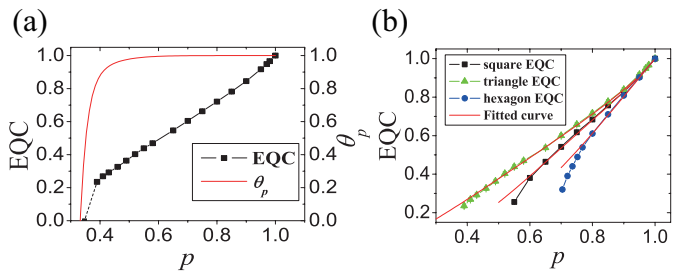}
\caption{\textbf{(a)} Comparison between EQC and $\theta_p$ of triangle lattice(the proportion of nodes in the largest connected nodes) \cite{enhance,theta}.
We may observe that $\theta_p$ (red linep) corresponding to entanglement percolation is
like a `jump function' demonstrating only success or failure,
while the behavior of the proposed EQC is different,
it in general increases monotonically with $p$ and approaches 1 finally when $p$ takes the limit of 1. Because the exact expression of $\theta_p$ is an infinite series and we picks first 30 terms here, the zero point of the curve is slightly different from the exact percolation threshold $0.347$ \cite{QEP}. But the `jump function' property is maintained. \textbf{(b)} Lines with
different symbols represent Monte Carlo estimation of EQC of square, triangle and hexagon lattices. Red lines are fitted curves according to Eq.\ p(\ref{fitform}), with corresponding lattice indices $\alpha_{square}\approx2.6$, $\alpha_{triangle}\approx0.9$, $\alpha_{hexagon}\approx2.8$.}
\label{Fit}
\end{figure}

Here we still consider the case of square lattice firstly. For the first situation (i), when $p>0.6$ which means $r_{at}<\sqrt {2}$,
the effective circle contains only 5 nodes, 4 internal bonds and 12 peripheral bonds connecting to the outside of the circle.
When Alice and Bob are far away from each other, the quantum channels
between $A$ and $B$ depend mostly on number of maximally entangled states
induced by those 4 internal bonds for Alice or Bob.
Only pairs of internal bonds belonging respectively to two sides
can possibly form independent quantum channels. The number of established pairs is denoted by $X$ which
is a random number. Thus the expectation of the number of pairs is,

\begin{equation}
\mathrm{E}X(p)=\sum_{1\leq i,j\leq 4}\min\{i,j\}C^i_{4}C^j_{4}p^{i+j}(1-p)^{8-(i+j)}.
\label{exppairs}
\end{equation}
We remark that when $\min\{i,j\}$ is omitted, the above equation equals to $4p$.
When $p\rightarrow1$, the medium can be seen as fully connected, and we have an asymptotic limit,
$E_0\approx {\mathrm{E}X(p)}/{4}$.

For general $p>0.6$, notice that for a connected neighboring point of $A$,
the probability that at least one peripheral bond is connected to the outside is $(1-(1-p)^3)$.
Under this consideration, the probability that the quantum connection does not meet a dead end in the `medium' can be written as $(1-(1-p)^3)^{\alpha}$,
where $\alpha$ is like an index of the medium depending only on lattice
which can be fitted by numerical data.
In this way, an estimated EQC can be written as follows,
\begin{equation}
E_0=\frac{\mathrm{E}X(p)}{4}(1-(1-p)^3)^{\alpha}
\label{fitform}
\end{equation}
We can find that this formula agrees well with the data, see FIG. \ref{Fit}.
The FIG. \ref{Fit}(a) also shows difference between entanglement percolation and EQC.
It is worth mentioning that the EQC also experienced a phase transition at the percolation threshold.

Next, we consider situation (ii).
Alice can access $k$ independent points $A_1$, $A_2$,..., $A_k$, Bob can similarly
access $B_1$, $B_2$,..., $B_k$. The same reasoning can be applied,
the Eq.(\ref{exppairs}) can be generalized as,
\begin{equation}
\mathrm{E}X(p)=\sum_{1\leq i,j\leq 4k}\min\{i,j\}C^i_{4k}C^j_{4k}p^{i+j}(1-p)^{8k-(i+j)}
\end{equation}
Then we immediately have,
\begin{equation}
E_0=\frac{\mathrm{E}X(p)}{4k}(1-(1-p)^3)^{\alpha}.
\label{kfit}
\end{equation}
The validity of this formula can be confirmed similarly as Eq.\ (\ref{fitform}),
and it agrees well with numerical results.
We remark that here we assume $p$ is far from the critical point.
Higher order of correction terms become
significant and thus not negligible when the considered $p$ gets closer to the critical point.
Actually the number of correction terms tend to infinity in Eq.\ (5) and Eq.\ (7)
when the estimation is carried out at the critical point of the network.

For comparison between Eq.\ (\ref{kfit}) and Monte Carlo approach, see \cite{appendix} for figures.
Data shows that multipartite-to-multipartite quantum communication
offers a higher EQC.
This can be explained as that more bonds tend to be easier to be paired up,
and the influence of the minimum function is weakened.

Another important kind of situation is that Alice and Bob can access different number of nodes. For example, Alice can access one node while Bob can access $k\geq$2 nodes. This resembles a `distribution' scenario. Results show that EQC is very close to $p$ and is almost constant when $k\geq2$. This can be explained similar to the previous case by noting that the four bonds within the attractive effective circle of Alice can almost always be paired in this case,
see \cite{appendix} for figures.

\section{Lattice Transformation}
Lattice transformation by entanglement swapping at some
nodes can enhance entanglement percolation properties \cite{QEP,enhance}. 
This is done by altering the lattice structure by carrying out first at some nodes
the optimal entanglement swapping protocol
and correspondingly changing
the percolation threshold of the network.
In this section we focus on the EQC of
the networks undertaking various transformations and study whether the lattice transformations can increase
or decrease the communication efficiency.
For lattice transformation, we need to modify the definition of EQC as
\begin{equation}
\mathrm{EQC}=\frac{N'(p)}{N(1)},
\end{equation}
where $N'(p)$ refers to the expectation number of quantum channels connecting
$A$ and $B$ in the new network, and $N(1)$ is that of the old network with $p=1$, regarded as fixed initial entanglement resources. (If two or more bonds are connecting two neighboring nodes, consider the bond-separated situations).
This modification takes lattice transformation effects into account, and the EQC can be compared to the previously defined
one to evaluate the effect of the transformations.

\begin{figure*}[lr]
\includegraphics[width=180mm]{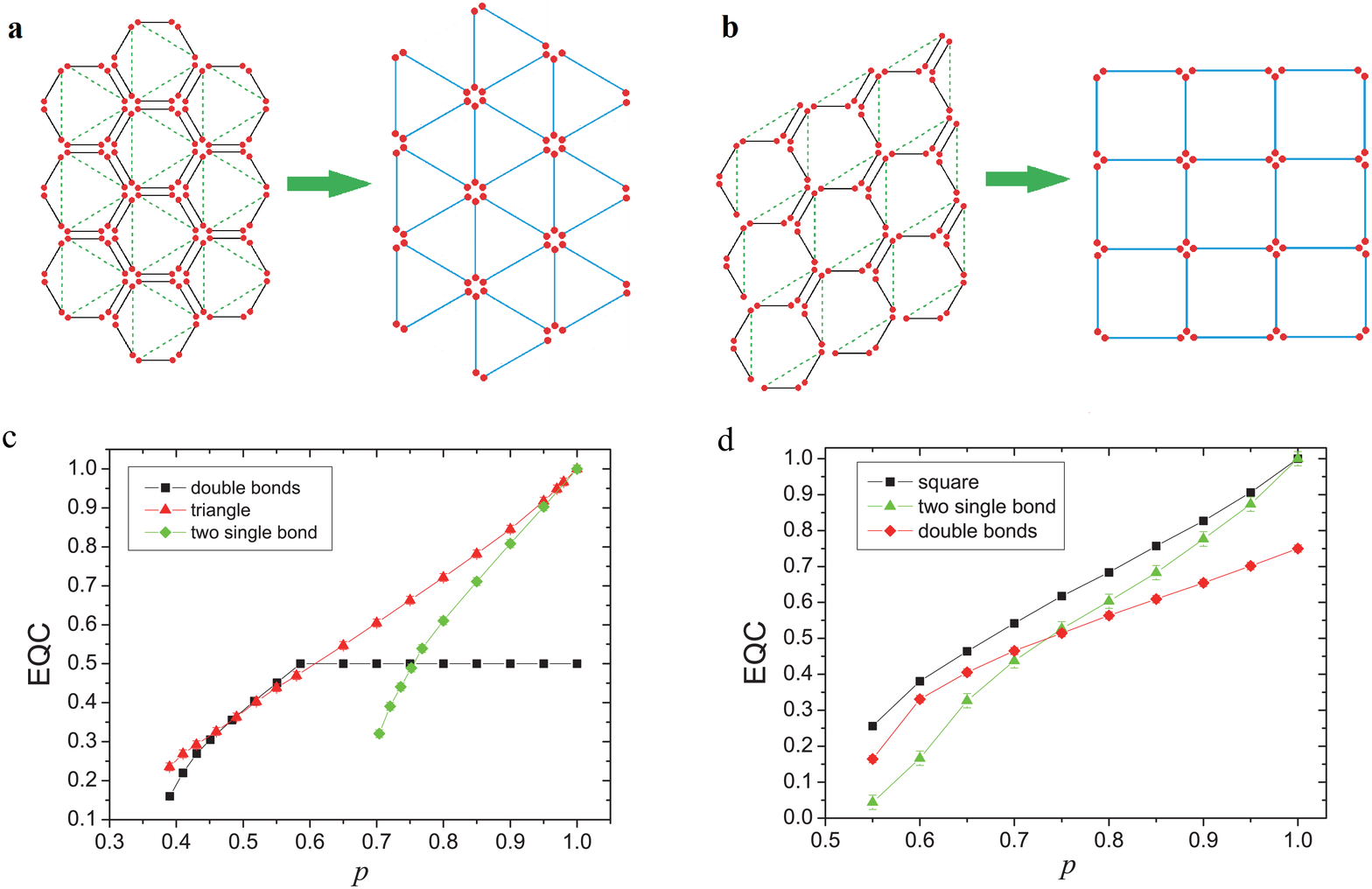}
\caption{\textbf{(a)} Double-bond hexagonal lattice and triangle lattice transformation via entanglement swapping.
\textbf{(b)} Hexagonal lattice with $1/3$ double bond and square lattice transformation via entanglement swapping. \textbf{(c)} Comparison between EQC of double-bond hexagonal lattice (black dotted line), single-bond triangular lattice (red dotted line) and two copies of single-bond hexagonal lattice (green dotted line). We can see that roughly in $0.5\leq p\leq 0.6$, the lattice transformation is not beneficial. In the limit $p\rightarrow 1$, the triangle lattice and two single-bond give almost the same EQC. \textbf{(d)} Comparison between EQC of hexagonal lattice with $1/3$ double bond (red dotted line), square lattice (black dotted line) and the hexagonal lattice regarding double bonds as two single bonds (green dotted line). We can see that after lattice transformation, not only the percolation threshold is decreased but also the EQC of square lattice for all $p$ is higher than those two different transformations about the double bonds. }
\label{LatticeTransform}
\end{figure*}

By using Monte Carlo methods, we observe two typical situations
when lattice transformations may be beneficial.
When the entanglement percolation have different thresholds
for the corresponding lattices, however, the EQC curves
may demonstrate two behaviors, \textbf{(i)} intersecting or \textbf{(ii)} non-intersecting.
For the \textbf{(i)} non-intersecting case, lattice transformation can enhance the efficiency
for all $p$ above the percolation threshold. In contrast for the \textbf{(ii)} intersecting case,
the lattice transformation induced by entanglement swapping is beneficial in some regions
while in other regions is not, i.e. the EQC may decrease after lattice transformation.
In the following, we consider two lattice transformations as examples.

\emph{Double-bond Hexagonal Lattice.}---
We know that a double-bond hexagonal lattice with $p$ slightly less than its percolation
threshold $p_{th6}$ can be transformed into a single-bond triangular lattice with $p'$ greater than
its percolation threshold $p_{th3}$ via entanglement swapping \cite{QEP,enhance}.
Our approach by estimating EQC reveals that there is an intersection between two EQC curves.
The corresponding $p$ of the intersection can be estimated by rewriting Eq.(\ref{fitform}) for triangle and hexagonal lattices $\mathrm{EQC}_{tri}=\mathrm{E}X(p)_{tri}(1-(1-p)^5)^{\alpha_{tri}}/6$,
$\mathrm{EQC}_{hex}=\mathrm{E}X(p')_{hex}(1-(1-p')^2)^{\alpha_{hex}}/6$.
We can substitute $p=2-\sqrt{2}$ and $p'=2(1-(1-p/2)^2)$  into those equations.
It is apparently that $EQC_{hex}=0.5$, while applying Eq. (\ref{fitform}) to triangle lattice,
we find $\mathrm{EQC}_{tri}<\mathrm{E}X(p)_{tri}/6=0.47<0.5$,
which demonstrates an intersection in the curves of EQC (see FIG. \ref{LatticeTransform}).

We also notice that for a network with
more bonds connecting with one node, its EQC is closer to Eq.(5).

Monte Carlo estimations confirm the intersection. Our approach provides an explicit illustration for
this lattice transformation: For $p\leq p_{intersect}$, lattice transformation is beneficial
not only for a lower critical point in entanglement percolation
but a higher efficiency shown by EQC as well.
But for $p\geq p_{intersect}$, transformation will decrease the quantum communication
efficiency represented by EQC.

Another lattice transformation involves separating double bonds and turns the network into two identical single-bond networks.
This strategy can be applied for around $p>0.76$ by considering the EQC (Fig.\ \ref{LatticeTransform}),
in which two separately-used bonds can give a higher $p$.

\emph{Hexagonal Lattice with $1/3$ Double Bonds to Square Lattice.}---
For this example, a hexagonal lattice with $\frac{1}{3}$ double bond can be transformed into a
single-bond square lattice (FIG. \ref{LatticeTransform}). First we can easily conclude that the percolation threshold of square lattice is lower than that of both kinds of strategies of hexagonal lattice with $1/3$ double bonds. Monte Carlo estimations show that the EQC of square lattice is higher than those two different strategies about the double bonds. This means we can benefit from lattice transformation for the whole range of $p>0.5$.

\section{Discussion}
Quantum network is currently one of the central topics of quantum information processing.
Quantum communication by teleportation protocol should be one fundamental function
of quantum networks where nodes share entanglements such as Einstein-Podolsky-Rosen pairs
as quantum channels. However, despite its many advantages, teleportation is entanglement resource
consuming which results in switching off the quantum channel when one state is teleported.
The number of independent or exclusive quantum channels shared by nodes of the quantum networks,
which corresponds the amount of quantum information capable of being transmitted,
becomes an important problem.
In this work, we systematically investigate this quantity and find some significant properties
by both analytical and numerical methods.
Quantum communication capacity quantified by EQC between nodes of quantum networks is, surprisingly,
independent of the communicating distance. This not only avoids the exponential decaying of
standard quantum repeating communication, but also it actually does not suffer any decaying
with distance. We remark that this non-decaying property is different from the
long-distance entanglement by that this quantum channel is established by a chain of
entanglement swapping. It naturally decays exponentially in one-dimensional case depending
on the conversion probability $p$.
Exact formulae of EQC are obtained, and the involved parameters are studied.
We also show that EQC can be enhanced by lattices transformations performed by quantum measurements.
At the same time, however, this may also induce the reduction of EQC in some region of $p$.
EQC is a new concept proposed specifically for quantum networks. Many aspects of this quantity
and some closely related topics can be explored further. This opens a new avenue in studying quantum
communication of quantum networks.

\emph{Acknowledgements:}
This research is supported by `973' program (2010CB922904), NSFC, China Innovation Funding
and grants from CAS.
\begin{widetext}
\section{Supplementary Information}
\section{Explicit curves of EQC}
See Fig. \ref{abcd}.
\begin{figure}
  \includegraphics[width=175mm]{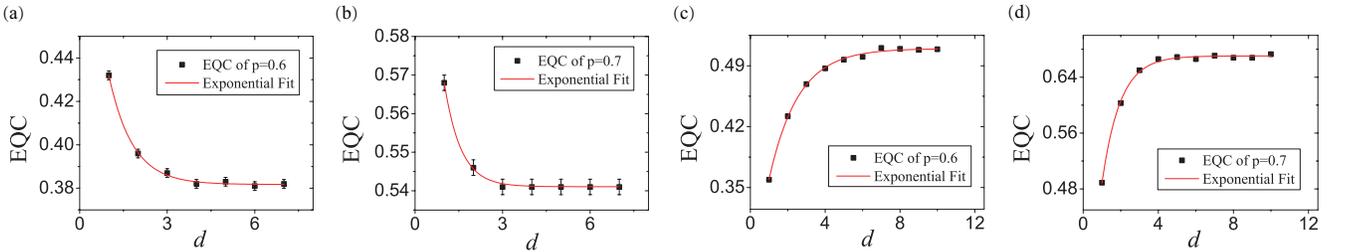}
  \caption{According to Eq. (2) and Eq. (3) in the main text, $E_0$ and $r_{at}$ ($r_{rp}$)are two important parameters of given network and approach of information transmission}. \textbf{(a)} $p=0.6$, the attractive radius $r_{at}\approx1.33$, the long distance EQC limit $E_0\approx0.38$. \textbf{(b)} $p=0.7$, the attractive radius $r_{at}\approx1.09$, the long distance EQC limit $E_0\approx0.54$. \textbf{(c)} $p=0.6$, the repulsive radius $r_{rp}\approx2.06$, the long distance EQC limit $E_0\approx0.5$. \textbf{(d)} $p=0.7$, the repulsive radius $r_{at}\approx1.46$, the long distance EQC limit $E_0\approx0.66$
  \label{abcd}
\end{figure}
\section{Comparison between Monte Carlo method and estimation Eq. (7)}
In our consideration in the main text, the probability of a successfully formed internal bond of sender is $p$. If all these bonds can finally connected to receiver, this gives the upper bond of EQC, that is, $EQC_{max}=p$. We can make comparison of EQCs between different situations and investigate how much are they closed to $EQC=p$. See Fig. \ref{e}
\begin{figure}
  \includegraphics[width=90mm]{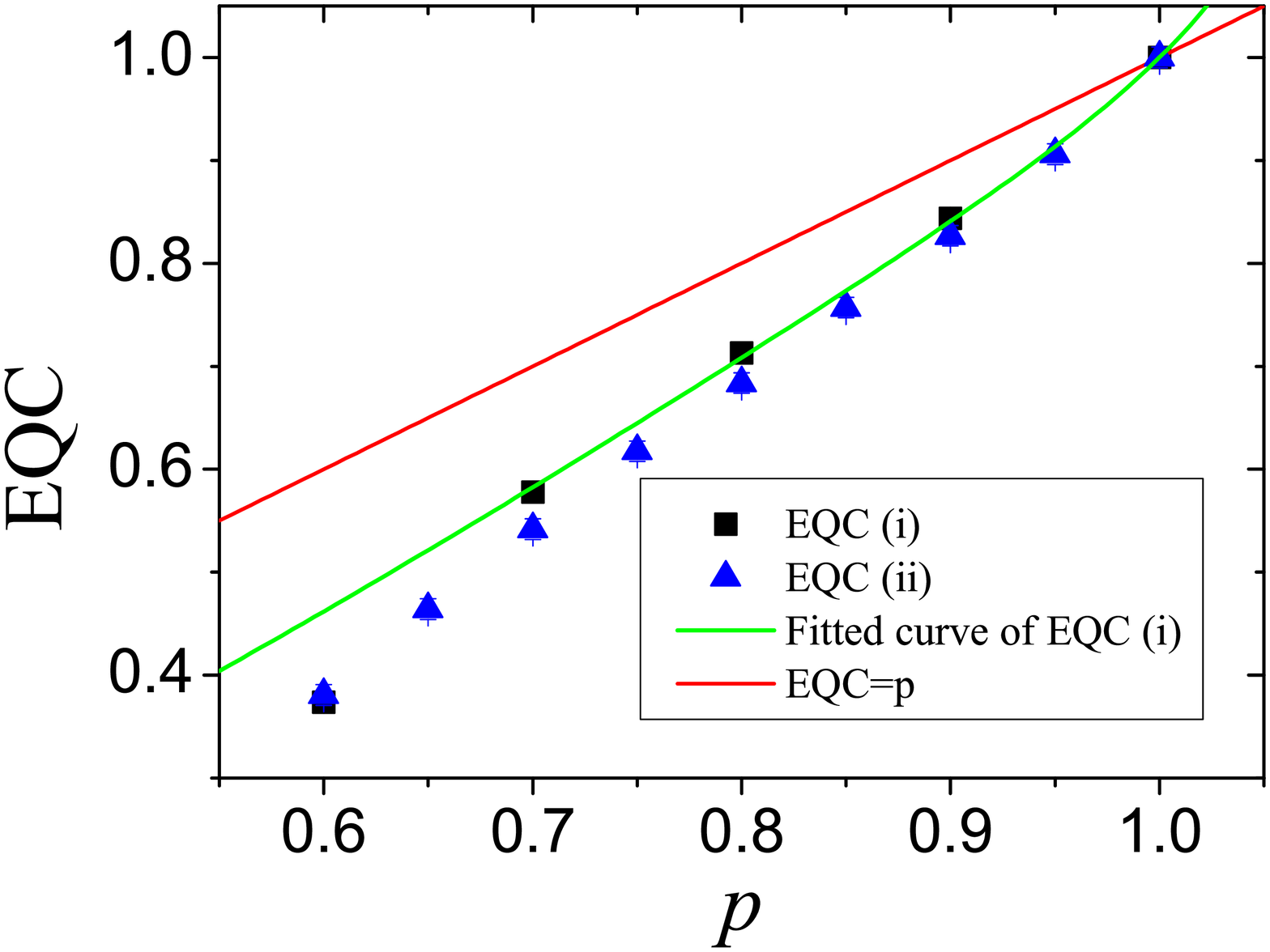}
  \caption{We show comparison of two scenarios. \textbf{(i)} In the main text, Eq. (7) gives the estimation of EQC $(E_0')$ of scenario where Alice and Bob both can access to $k$ independent nodes. Black squares represent EQC (i) with $k=2$. \textbf{(ii)} Alice and Bob both can access one nodes. Blue triangles represent EQC (ii). Green line is the estimation function Eq. (7). Red line denotes $EQC=p$, which is the maximum number of long-distance EQC according to Eq.(4)-(7) in the main text}.
  \label{e}
\end{figure}
\section{Situation when Alice and Bob can access to different number of nodes}
According to Eq. (6) in the main text, the influence of the minimum function will be reduced, thus gives a higher EQC, when Alice and Bob can access different number of nodes. It is because if the number of nodes accessed by receiver is larger than that of sender, the internal bonds of the sender can almost always be paired by internal bonds of receiver's. See Fig. \ref{f}
\begin{figure}
  \includegraphics[width=90mm]{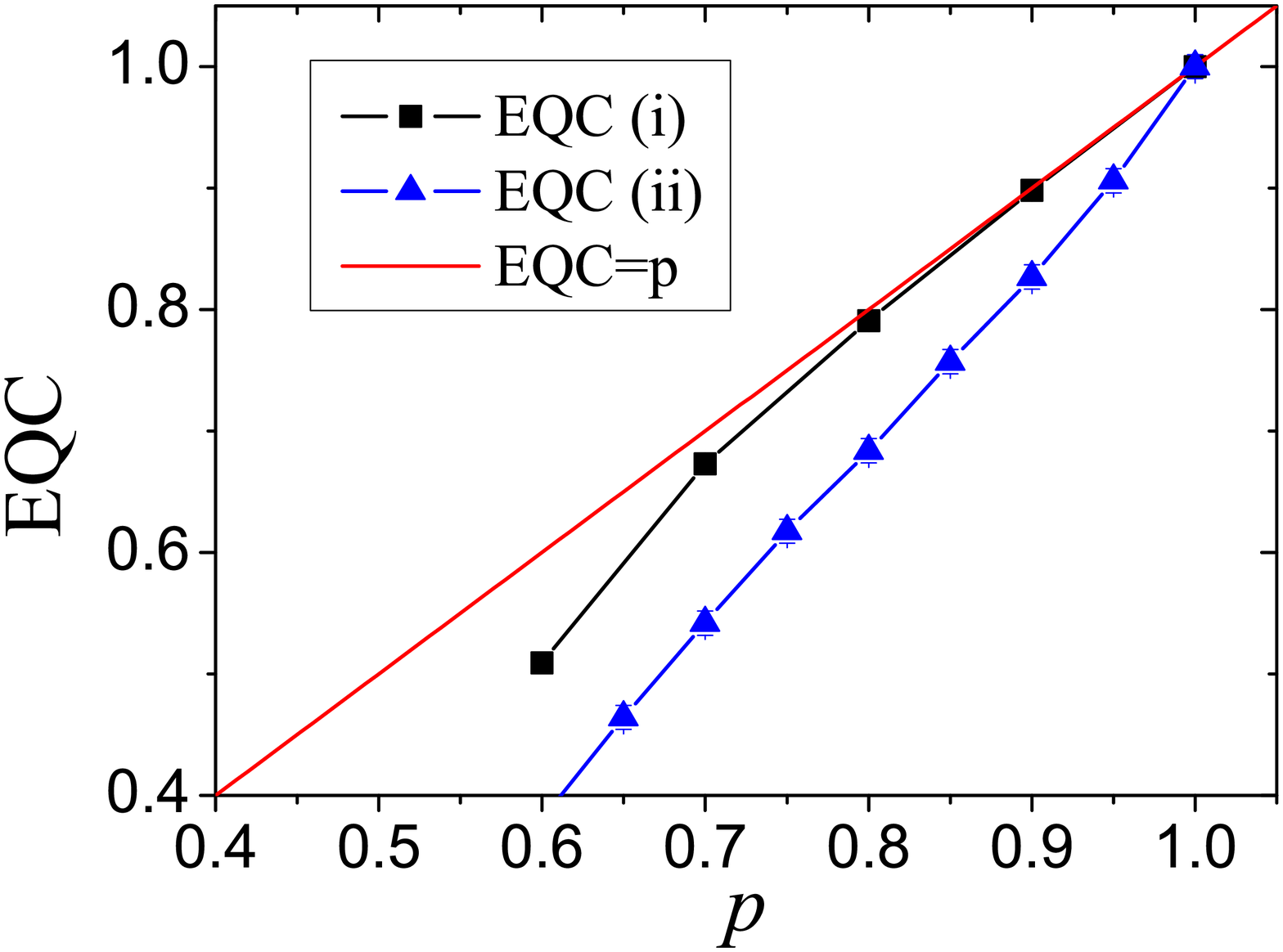}
  \caption{Black squares and lines represents the `distribution' scenario related to the repulsive radius in the main text: Alice and Bob, each accesses to one node, send qubits to many receivers far away. According to the reasoning below, these `distribution' scenarios have higher EQC, thus higher efficiency. We can observe this efficiency shows asymptotic behavior of $EQC=p$. Blue triangles and lines represents scenario where Alice and Bob both can access one node,  sending qubits to each other. The EQC of this type is much lower.}
  \label{f}
\end{figure}
\end{widetext}

\end{document}